\documentclass[twocolumn,eqsecnum,showpacs,preprintnumbers,amsmath,amssymb,prd]{revtex4}
\usepackage{natbib}
\usepackage{amsmath}
\usepackage{amssymb}
\usepackage{graphicx}

\def\ts{\textstyle}

\def\be{\begin{equation}}
\def\ee{\end{equation}}
\def\nbar{$\bar{n}\;$}
\def\ba{\begin{eqnarray}}
\def\ea{\end{eqnarray}}
\def\gal{galaxies/arcmin$^2\;$}
\def\zmax{$z_{max}\;$}

\newcommand{\mx}{\mbox}
\newcommand{\bm}{\boldmath }
\newcommand{\nn}{\nonumber \\}
\newcommand{\ax}{$\approx$}
\newcommand{\st}{$\leq$}
\newcommand{\gt}{$\geq$}

\def\apjl{Astrophys.\ J.\ Lett.}    

  \def\aap{Astron.\ Astrophys.}
\def\apj{Astrophys.\ J.}    \def\apjs{Astrophys.\ J. Supp.}  
  \def\prd{Phys.\ Rev.\ D}
  \def\physrep{Phys. Rep.}
\begin{document}
\title{Detectability of CMB tensor $B$ modes via delensing with weak
lensing galaxy surveys}

\author{Laura Marian and Gary M. Bernstein \vspace{0.2cm}}

\affiliation{Department of Physics and Astronomy, University of Pennsylvania, Philadelphia, Pennsylvania 19104, USA}
\date{July 2, 2007}

\begin{abstract}
We analyze the possibility of delensing Cosmic Microwave Background
(CMB) polarization maps using foreground weak lensing (WL)
information. We build an estimator of the CMB lensing potential out of
optimally combined projected potential estimators to different source
redshift bins. Our estimator is most sensitive to the redshift depth
of the WL survey, less so to the shape noise level. Estimators built
using galaxy surveys like LSST and SNAP recover up to 80-90\% of the
potential fluctuations power at $l\leq 100$ but only \ax 10-20\% of
the small-angular-scale power ($l\leq 1000$). This translates into a
30-50\% reduction in the lensing $B$-mode power. \par We illustrate
the potential advantages of a 21-cm survey by considering a fiducial
WL survey for which we take the redshift depth \zmax and the effective
angular concentration of sources \nbar as free parameters. For a noise
level of 1 $\mu$K arcmin in the polarization map itself, as projected
for a CMBPol experiment, and a beam with $\theta_{\rm\scriptscriptstyle{FWHM}}$=10
arcmin, we find that going to \zmax=20 at \nbar=100 \gal yields a
delensing performance similar to that of a quadratic lensing potential
estimator applied to small-scale CMB maps: the lensing $B$-mode
contamination is reduced by almost an order of magnitude. In this
case, there is also a reduction by a factor of \ax4 in the
detectability threshold of the tensor $B$-mode power. At this CMB
noise level, the $B$-mode detection threshold is only $3\times$ lower
even for perfect delensing, so there is little gain from sources with
$z_{max}>20$. The delensing gains are lost if the CMB beam exceeds
$\sim 20$ arcmin.  The delensing gains and useful $z_{max}$ depend
acutely on the CMB map noise level, but beam sizes below 10 arcmin do
not help. Delensing via foreground sources does not require
arcminute-resolution CMB observations, a substantial practical
advantage over the use of CMB observables for delensing.
\end{abstract}
%
\maketitle

\section{Introduction}
\label{I}
The anisotropies in the CMB have been long recognized as a major probe
for cosmology. The WMAP satellite has measured the temperature
fluctuations up to $l_{max} <=$1000 and has confirmed the cosmological
standard model of a power-law, flat, $\Lambda$CDM universe.  Much hope
lies with CMB polarization measurements because they have the
potential to unveil some of the unknowns of inflation. While some
predictions of inflation (such as nearly flat space curvature, nearly
scale-invariant power spectrum, and Gaussianity of the primordial
fluctuations) have been confirmed by CMB and large scale structure
experiments, there is another prediction which is yet to be
verified. This is the existence of an almost scale-invariant spectrum
of gravitational waves, whose amplitude is directly related to the
energy scale of inflation. The inflationary gravitational waves are
tensor perturbations to the metric and we expect them to leave a
curl-like signature in the CMB polarization field, i.e. a $B$-mode
pattern. Density fluctuations, arising from scalar perturbations to
the metric, create a gradient-like component in the polarization
field, the $E$ mode. In the linear regime, density fluctuations do not
create a $B$ mode, so a detection of the latter would confirm the
existence of primordial gravitational waves; it would also provide the
energy scale of inflation, which we could use to distinguish between
different inflationary scenarios.\par CMB polarization measurements
are difficult to carry out, primarily for two reasons. Firstly, the
amplitude of the signal is very small: for example, the scalar
$E$-mode power is 2-3 orders of magnitude smaller than the scalar
temperature power, tensor $B$-mode much lower. Secondly, the
polarization foregrounds are very poorly understood and they dominate
the CMB signal at almost all relevant frequencies. Therefore,
foreground removal and detector sensitivity are two of the most
stifling limitations of a polarization experiment.\par $B$-mode
measurements are additionally obstructed by WL contamination,
especially of the recombination signal (l$\leq$100). Thus CMB
polarization measurements are able to provide inflationary insights to
the extent to which we can: 1. remove the foreground contribution to
the overall signal and 2. delens the $B$ mode power and extract the
tensor contribution to it. The delensing process consists of the
reconstruction of the lensing potential from chosen observables. The
estimated lensing potential is then used to evaluate the WL-created
$B$-mode signal and to subtract it from the measured $B$-mode map.
\par In this paper we probe the ability of weak lensing (WL) galaxy
surveys to delens the CMB in the absence of foregrounds. To be
specific, we try to answer two questions: what attributes should a
galaxy or 21~cm WL survey and also a CMB polarization mission have in
order to detect the tensor $B$ mode? What is the minimum amplitude of
the $B$ mode, expressed in terms of the tensor-to-scalar ratio, $r$,
that can be detected using galaxy or recombination observations for
delensing?  We take three examples of surveys to illustrate how our
estimator works: the ground-based Large Synoptic Survey Telescope
(LSST \footnote{\tt {http://www.lsst.org/}}), the space-based
Supernova Acceleration Probe (SNAP \footnote{\tt
{http://snap.lbl.gov/}}) and a toy model mimicking recombination-era
21-cm observations that we mention in more detail in section
\S\ref{IV}. The outline of the paper is as follows.
In \S\ref{II} we describe the WL contamination of the tensor $B$
mode. In \S\ref{III} we present our minimum-variance lensing
estimator. In \S\ref{IV} we determine the minimum detectable $r$ when
we use this estimator to delens.  In \S\ref{V} we discuss our results
and draw conclusions. Let us now briefly mention similar work existing
in the literature.  There has been a vast and impressive amount of
work on the topic of CMB delensing and $r$-detection. The great
majority of this work uses the CMB observables $\Theta, E, B$ to
reconstruct the projected potential. Averaging over various quadratic
combinations of the temperature field (e.g. see the work of
\citet{1999PhRvD..59l3507Z}, \citet{1998A&A...338..767B},
\citet{2001PhRvD..64h3005H}, \citet{2001ApJ...557L..79H}) and the
polarization field (e.g. \citet{2000PhRvD..62d3517G},
\citet{2002ApJ...574..566H}, \citet{2003PhRvD..67l3507K}) has been
thoroughly considered for the projected potential reconstruction. To
give a quick summary: one can build minimum-variance, unbiased
estimators, using certain field statistics, as shown by
\citet{2001ApJ...557L..79H}. For a post-Planck experiment (sensitivity
of 0.3 $\mu$K arcmin and beam size of 3 arcmin),
\citet{2002ApJ...574..566H} found that the most efficient of these
estimators can map the potential up to
l$\leq$1000. \citet{2002PhRvL..89a1304K} and
\citet{2002PhRvL..89a1303K} used this last estimator to predict the
minimum detectable $r$ as a function of CMB experimental
characteristics. There is another, more promising method for lensing
potential reconstruction, based on likelihood techniques.
\citet{2003PhRvD..67d3001H} have built a maximum-likelihood estimator
for the convergence field using temperature maps and have found its
performance similar to that of the quadratic estimator introduced by
\citet{2001ApJ...557L..79H}. \citet{2003PhRvD..68h3002H} found that
the same maximum likelihood estimator built from polarization maps is
even more effective: there is an order of magnitude reduction in the
mean squared error in the lensing reconstruction compared to the
quadratic estimator method, if the survey characteristics are
adequate: sensitivity of 0.25 $\mu$K arcmin and a beam size of 2
arcmin. \citet{2005PhRvD..72l3006A} analyze the detectability of
tensor $B$ modes in the presence of polarized dust emission, as a
function of sky coverage; \citet{2006JCAP...01..019V} do a study of
optimal surveys for $B$-mode detection, considering both dust and
synchrotron emissions.  One disadvantage of the reconstruction methods
presented so far is that they require high-resolution CMB maps: they
use arc-minute structures of the CMB fields, to reconstruct
degree-scale maps of the deflection field, as explained by
\citet{2001ApJ...557L..79H}. \citet{2005PhRvL..95u1303S} point out
that one can use non-CMB observables to delens the CMB; in this case
the requirement for high angular resolution of the CMB mission can be
relaxed significantly. These authors determine lower limits for the
detectable $r$ using the 21 cm radiation emitted by neutral hydrogen
atoms to delens the CMB. We follow a similar approach here, but employ
foreground galaxies instead of 21-cm emission as the source plane for
delensing.
\section{B modes and Weak Lensing}
\label{II}
\par As mentioned in \S\ref{I}, inflation predicts scalar and tensor
perturbations; the former are responsible for density fluctuations and
for the formation of structure. The latter are gravitational
waves. Direct detection of gravitational waves has been a physicists'
dream for a long time and there is ongoing effort to make this dream
come true (e.g. see LISA \footnote{\tt{http://lisa.nasa.gov/}}, LIGO
\footnote{\tt{http://www.ligo.caltech.edu/}}). There is however another
means to ascertain the existence of tensor perturbations, by measuring
the CMB polarization.\par The $E$ and $B$ modes are defined as
linear combinations of the Stokes parameters $Q$ and $U$, such that
the power of $B$ arising from linear order $\emph{scalar
perturbations}$ is zero, as shown by
\citet{1997PhRvD..55.1830Z}.
Tensor perturbations yield both $E$ and $B$
modes, roughly of the same magnitude; both modes could in principle
provide information on gravitational waves and inflation, but in
practice the tensor $E$ modes are overwhelmed by their
scalar counterparts. 
The amplitude of tensor modes is quantified by the tensor-to-scalar
ratio $r,$ the ratio of the expectation values for the quadrupole
tensor and scalar temperature anisotropies.  \par Even if one assumes
flawless detector technology, $B$-mode measurements are plagued by the
presence of polarized foregrounds and by WL contamination. A detailed
discussion of the foreground problem is beyond the scope of this work
and we refer the interested reader to the Final Report of the Task
Force on CMB Research \citep{2006astro.ph..4101B}. We briefly mention
that while most galactic polarized foreground emission is currently
not well known at the frequencies of interest to CMB experiments,
there is reason to believe that this situation will change in the near
future. Indeed, the WMAP satellite is gathering all-sky data on
the synchrotron emission and the HFI instrument on the Planck
satellite will map the dust emission. Numerous smaller, ground-based
missions, with higher angular resolution, plan to complement the
above-mentioned satellite data. All this information should
enable the study of polarized foreground emission as a function of
both frequency and angular scale.  \par We now sketch the equations
that describe the WL contamination of the tensor $B$ mode. The groundwork
on this subject was done by \citet{1998PhRvD..58b3003Z}; for a more
recent review see \citet{2006PhR...429....1L}. A photon travelling
from the last scattering surface is affected by potential wells along
the line of sight. Its frequency changes because the potentials vary
with time (the Rees-Sciama effect, the integrated Sachs-Wolfe effect)
and its path can be transversally shifted by lensing. To a good
approximation, the lensed wave has the same Stokes parameters along a
path $\mx{\bm$\hat n$}$ as the unlensed one, only shifted by
the average deflection along that path:
$$\left( \begin{array}{c}
\tilde{I}\\
\tilde{Q}\\
\tilde{U}\\
\end{array}
\right)\left(\mx{\bm$\hat{n}$}\right)=\left( \begin{array}{c}
I\\
Q\\
U\\
\end{array}
\right)\left(\mx{\bm$\hat{n}$}+ \mx{\bm$\alpha$} \right).
$$
where we denote by $\tilde{X}$ the lensed field of parameter $X$, and 
 $\mx{\bm$\alpha$}$ is the deflection angle. To
linear order, the deflection angle is the gradient of the projected
potential: $\mx{\bm$\alpha$=$\nabla$}\psi$. Assuming that
deflection angles are very small compared to the characteristic scale
of fluctuations, we can Taylor expand the lensed parameters around the
undeflected path $\hat x$. Using the flat-sky approximation, we can
write the lensed $B$ mode in Fourier space as:
\ba
\tilde B(\mx{\bm$l$})=B(\mx{\bm$l$})-\int\frac{d^{2}l'}{2\pi}f\left(\mx{\bm$l'$,\,\bm$l$}\right)\,\psi(\mx{\bm$L$})E(\mx{\bm$l'$})-\hspace{0.5in} & \nonumber\\
-\frac{1}{2}\int\frac{d^{2}l'}{2\pi}\frac{d^{2}l''}{2\pi}g\left(\mx{\bm$l'$,\,\bm$l''$,\,\bm$l$}\right)\psi(\mx{\bm$l''$})\psi^{*}(\mx{\bm$l''$}\!\!-\mx{\bm$L$})E(\mx{\bm$l'$}).\hspace{0.1in}
\label{eq:lensedB}                                         
\ea
Here $B$ denotes the tensor $B$ mode and $E$ is the unlensed, scalar $E$
mode. The above equation neglects the lensing
effect on the tensor $E$ and $B$ fields, since it is very small
compared to the lensing of the scalar $E$ mode. The functions $f$ and $g$ are
defined by:
\ba
f\left(\mx{\bm$l'$,\,\bm$l$}\right)=\mx{\bm$l'$}\cdot\mx{\bm$L$}\,\sin[2(\phi_{l'}-\phi_{l})]\hspace{1.25in}  & \nonumber\\
g\left(\mx{\bm$l'$,\,\bm$l''$,\,\bm$l$}\right)=\mx{\bm$l'$}\cdot (\mx{\bm$l''$}\!\!-\mx{\bm$L$})(\mx{\bm$l'$}\cdot \mx{\bm$l''$})\,\sin[2(\phi_{l'}-\phi_{l})]\,,\hspace{0.4in}
\ea
where $\mx{\bm$L$}=\mx{\bm$l$}-\mx{\bm$l'$}$
and $\phi_{l}$ is the angle between $\mx{\bm$l$}$ and the
$\mx{\bm$x$}$ axis.  We assume statistical isotropy and parity
invariance and we adopt the power spectrum definition of
\citet{2006PhR...429....1L}:
\ba
\langle E(\mx{\bm$l$})E^{*}(\mx{\bm$l'$})\rangle&=&\delta(\mx{\bm$l$}-\mx{\bm$l'$})C_{l}^{E}, \nonumber\\
\langle B(\mx{\bm$l$})B^{*}(\mx{\bm$l'$})\rangle&=&\delta(\mx{\bm$l$}-\mx{\bm$l'$})C_{l}^{B}, \nonumber \\
\langle B(\mx{\bm$l$})E^{*}(\mx{\bm$l$})\rangle&=&0.
\label{eq:PowerDef}
\ea
With the above considerations in mind, we write down the lensed B-mode
power spectrum as a sum of the tensor (primordial) $B$-mode power and
the weak lensing $B$-mode power:
\be
\tilde C^{B}_{l}=C^{B}_{l}+C_{l}^{\scriptscriptstyle WL},
\label{eq:lensedClB1}
\end{equation}
with the appropriate definitions:
\ba
C^{B}_{l}=r\,C_{l}^{tensor}\hspace{2.13in} \label{eq:trueClB}& \\
C^{{\scriptscriptstyle WL}}_{l}=\int\frac{d^{2}l'}{(2\pi)^{2}}\,|f\left(\mx{\bm$l'$,\,\bm$l$}\right)|^{2}C^{\psi}_{L}C^{E}_{l'}\,+ \hspace{1in} & \nonumber\\
+\frac{1}{4}\int\frac{d^{2}l'}{(2\pi)^{2}}\frac{d^{2}l''}{(2\pi)^{2}}\,|g\left(\mx{\bm$l'$,\,\bm$l''$,\,\bm$l$}\right)|^{2}\,C^{\psi}_{l''}\,C^{\psi}_{|\mx{\bm$\scriptstyle l'' -L $}|}\,C^{E}_{l'}\, , \hspace{0.25in}
\label{eq:lensedClB2}
\ea
where $C^{B}_{l}$ is the tensor $B$ mode power spectrum, $C^{E}_{l}$
 is the power spectrum of the unlensed scalar $E$ mode and
 $C^{\psi}_{l}$ is the lensing potential power spectrum.
\section{The CMB potential estimator}
\label{III}
\subsection{The ideal case}
\label{IIIa}
The measurable $B$-mode power is a sum of the primordial $B$-mode
tensor power and a convolution in Fourier space of the scalar $E$-mode
power with the lensing potential power, as shown in
Eq.~(\ref{eq:lensedClB1}). In order to have access to the tensor $B$
mode, we need to subtract the lensing contribution from the measured
$B$-mode map. The impediments are that we know neither the lensing
potential nor the unlensed $E$ mode power.  \par In this section we
address the first of these problems and build a lensing potential
estimator using WL foreground galaxy survey information, independent
of a CMB experiment. Unlike the estimators built out of CMB
observables, our estimator does not require extremely high resolution
in the CMB polarization map. This is an important advantage, since
high resolution is the main cost driver of CMB polarization
experiments. The disadvantage of our approach is that WL surveys
detect source galaxies up to a certain redshift, depending on the
specifics of the survey. Our estimator cannot reconstruct the
potential fluctuations evolving between that upper-limit redshift and
the redshift of recombination.  \par We build the lensing potential
estimator as a weighted sum of the projected potential measured to
different source galaxy redshift bins. We choose the weights so that
the estimator be optimal, i.e. the variance of the error in each mode
is minimal. We work in the flat sky approximation and we assume a flat
Friedmann-Robertson-Walker (FRW) metric. We write the estimator as:
\be
\hat\psi_{\scriptscriptstyle \rm{CMB}}(\mx{\bm$l$})=\sum_{i}\alpha_{i}(\mx{\bm$l$})\hat\psi_{i}(\mx{\bm$l$}),
\label{eq:raw_est}
\ee
where the sum is over source bins $i$ and the
$\mx{\bm$\alpha$}'s$ weight the contribution of the projected
potential to each redshift source bin. The projected potential to the
ith source bin is defined by:
\be
\psi_{i}(\mx{\bm$\theta$})=\int_{0}^{\chi_{\infty}}\!\! d\chi \, W_{i}(\chi)\Phi^{3D}(\chi \mx{\bm$\theta$},\chi),
\ee
where $\Phi^{3D}(\chi \mx{\bm$\theta$},\chi)$ is the 3D gravitational 
potential spectrum. In Eq.~(\ref{eq:raw_est}) we use the
Fourier transform of the projected potential:
$\psi_{i}(\mx{\bm$l$})=\int d^{2}\theta
e^{-\ts{i}\mx{\bm$l\cdot\theta$}}\psi_{i}(\mx{\bm$\theta$})$.
The lensing weight function $W_{i}(\chi)$ of the source bin $i$ is given by:
\be
W_{i}(\chi)=\frac{\ts 2}{\ts c^{2}}\frac{\ts{\int_{z_{i}}^{z_{i+1}}dz_{s}\mathcal{P}(z_{s})}\frac{\ts{\chi(z_{s})-\chi}}{\ts{\chi(z_{s})\,\chi}}\;\scriptstyle{\mathcal {U}}\ts{(z_{s}-z(\chi))}}{\ts{\int_{z_{i}}^{z_{i+1}}dz_{s}\mathcal{P}(z_{s})}}.
\label{eq:lwf}
\ee
Here $\scriptstyle{\mathcal {U}}\ts{(x)}$ is the Heaviside unit step function and $\mathcal{P}(z_{s})$ is the redshift distribution of the source
galaxies.
The residual of our estimator, defined as: 
$$\mathcal{R}^{\psi}(\mx{\bm$l$})\equiv \psi_{\scriptscriptstyle
  \rm{CMB}}(\mx{\bm$l$})-\hat\psi_{\scriptscriptstyle
  \rm{CMB}}(\mx{\bm$l$})\,,$$ tells us how the estimator
compares to the true CMB projected potential:
$$\psi_{\scriptscriptstyle \rm{CMB}}(\mx{\bm$\theta$})=\int_{0}^{\chi_{\infty}}\!\! d\chi\, W_{\scriptscriptstyle \rm{CMB}}(\chi)\Phi^{3D}(\chi \mx{\bm$\theta$},\chi),
$$
where $W_{\scriptscriptstyle \rm{CMB}}(\chi)$ is the same weight as in
eq.~(\ref{eq:lwf}), but written for the particular case of one source
at redshift $z_{\scriptscriptstyle \rm{CMB}}\,,\,$
i.e. $\mathcal{P}(z_{s})=\delta_{D}(z_{\scriptscriptstyle \rm{CMB}}-z_{s})$:
$$W_{\scriptscriptstyle \rm{CMB}}(\chi)=\frac{2}{c^{2}}\frac{\chi(z_{\scriptscriptstyle \rm{CMB}})-\chi}{\chi(z_{\scriptscriptstyle \rm{CMB}})\,\chi}.
$$
We determine the weights \mx{\bm$\alpha$} by requiring that the variance of the above-mentioned residual be minimal:
\be
\frac{\partial}{\partial \alpha_{k}(\mx{\bm$l$})}\langle\, |\psi_{\scriptscriptstyle \rm{CMB}}(\mx{\bm$l$})-\hat{\psi}_{\scriptscriptstyle \rm{CMB}}(\mx{\bm$l$})|^{2}\,\rangle=0\,,\; \forall k.
\label{eq:alpha1}
\ee
To see the solution of this equation, let us express the variance of the estimator's residual as a matrix equation:
\ba
\langle\, |\psi_{\scriptscriptstyle \rm{CMB}}(\mx{\bm$l$})-\hat{\psi}_{\scriptscriptstyle \rm{CMB}}(\mx{\bm$l$})|^{2}\,\rangle=\mx{\bm$\rm{\alpha}$}(\mx{\bm$l$})\cdot \mx{\bm$\rm{\mathcal {W}}$}^{\rm{g g}}(l)\cdot\mx{\bm$\rm{\alpha}$}^{t}(\mx{\bm$l$}) & \nonumber\\
-2\mx{\bm$\rm{\alpha}$}(\mx{\bm$l$})\cdot \mx{\bm$\rm{\mathcal {W}}$}^{\rm{g \scriptscriptstyle CMB}}(l)+\mx{\bm$\rm{\mathcal {W}}$}^{\rm{\scriptscriptstyle CMB\,\scriptscriptstyle CMB}}(l)\:.\hspace{0.8in}
\label{eq:VarRes1}
\ea
In the above equation we have used the following definitions:
\ba
\mx{\bm$\rm{\mathcal W}$}_{i\,j}^{\rm{g g}}(l) & \equiv & \langle\psi_{i}(\mx{\bm$l$})\,\psi_{j}^{*}(\mx{\bm$l$})\rangle,  \nonumber\\
\mx{\bm$\rm{\mathcal W}$}_{i}^{\rm{g \scriptscriptstyle{CMB}}}(l) & \equiv & \langle\psi_{i}(\mx{\bm$l$})\,\psi_{\rm{\scriptscriptstyle{CMB}}}^{*}(\mx{\bm$l$})\rangle,  \nonumber\\
\mx{\bm$\rm{\mathcal {W}^{\scriptscriptstyle{CMB}\,\scriptscriptstyle{CMB}}}$}(l) & \equiv &  \langle\psi_{\rm\scriptscriptstyle{CMB}}(\mx{\bm$l$})\,\psi_{\rm{\scriptscriptstyle{CMB}}}^{*}(\mx{\bm$l$})\rangle.
\ea
Using the relation $P_{\Phi}(k)=\left[\frac{3}{2}\,H_{0}^{2}\,\Omega_{m}^{0}\,(1+z)\right]^{2}\frac{\ts{1}}{\ts{k^{4}}}\,P_{\delta}(k)$,
to go from the potential power spectrum to the density power spectrum
and also the Limber approximation, we calculate the
\mx{\bm$\mathcal W$}-matrices:
\ba
\mx{\bm$\rm{\mathcal W}$}_{i\,j}^{\rm{g g}}(l)=\frac{\rm{\mathcal W_{0}}^{2}}{l^{4}}\!\!\int_{0}^{\infty}\!\!dz\frac{d\chi}{dz}(1+z)^{2}\,W_{i}(\chi)W_{j}(\chi) P_{\delta}(\frac{l}{\chi}),\hspace{0.29in} & \nonumber\\
\mx{\bm$\rm{\mathcal W}$}_{i}^{\rm{g \scriptscriptstyle{CMB}}}(l)=\frac{\rm{\mathcal W_{0}}^{2}}{l^{4}}\!\!\int_{0}^{\infty}\!\!dz\frac{d\chi}{dz}(1+z)^{2}W_{i}(\chi)W_{\scriptscriptstyle \rm{CMB}}(\chi)P_{\delta}(\frac{l}{\chi}),\hspace{0.0in} & \nonumber\\
\mx{\bm$\rm{\mathcal {W}^{\scriptscriptstyle{CMB}\,\scriptscriptstyle{CMB}}}$}(l)=\frac{\rm{\mathcal W_{0}}^{2}}{l^{4}}\!\!\int_{0}^{\infty}\!\!dz\frac{d\chi}{dz}(1+z)^{2}\,W_{\scriptscriptstyle \rm{CMB}}^{2}(\chi)P_{\delta}(\frac{l}{\chi})\,,\hspace{0.04in}&\nonumber\\
\label{eq:Wmatrix}
\ea
where $\rm{\mathcal W_{0}}\,$$\equiv 3H_{0}^{2}\,\Omega_{m}^{0}$

\par We are now able to write down the weights
\mx{\bm$\alpha$} that minimize the variance of our estimator's
residual:
\be
\mx{\bm$\rm{\alpha}$}(l)=\mx{\bm$\rm{\mathcal W} $}^{\rm{g \scriptscriptstyle CMB}}(l)\cdot \left(\mx{\bm$\rm{\mathcal W} $}^{\rm{g g}}\right)^{-1}(l).
\label{eq:alpha2}
\ee
We note that the weights depend only on the magnitude of
\mx{\bm$l$}. Our next goal is to rewrite the expression of
the variance of $\mathcal R(\mx{\bm$l$})$, using
eq.~(\ref{eq:alpha2}); the $l$-dependence is implicit here:
\be
\rm{Var}\,\mathcal{R}^{\psi}=\mx{\bm$\rm{\mathcal W}$}^{\rm{\scriptscriptstyle CMB\,\scriptscriptstyle CMB}}-\mx{\bm$\rm{\mathcal W}$}^{\rm{g \scriptscriptstyle CMB}}\cdot \left(\mx{\bm$\rm{\mathcal W}$}^{\rm{g g}}\right)^{-1}\cdot\mx{\bm$\rm{\mathcal W}$}^{\rm{g \scriptscriptstyle CMB}}.
\label{eq:VarR}
\ee
 The modes where ${\rm Var}{\mathcal R}(l)\rightarrow 0$ are well reconstructed by our estimator.
\subsection{The effect of Shape Noise}
\label{IIIb}
\par So far we have considered an ideal case, with clean WL
measurements. In fact we expect the shape noise of galaxies to limit
the performance of our estimator quite significantly. The intrinsic
ellipticities of galaxies increase the variance of the projected
potential in each source bin. The shape noise effect on
$\mx{\bm$\rm{\mathcal W}$}^{\rm{g g}}(l)$ is:
\be
\left(\mx{\bm$\rm{\mathcal W}$}_{i\,j}^{\rm{g g}}\right)_{SN}(l)=\mx{\bm$\rm{\mathcal W}$}_{i\,j}^{\rm{g g}}(l)+\frac{\sigma_{\gamma}^{2}}{\bar n_{i}}\,\frac{4}{l^{4}}\,\delta_{i\,j}\,,
\label{eq:W_SN}
\ee
where $\sigma_{\gamma}$ is the uncertainty in the measurement of one
galaxy and $\bar n_{i}$ is the number density of sources in bin $i$.
\subsection{The CMB Lensing Potential Estimator for Fiducial Surveys}
\label{IIIc}
\par We use three examples of WL surveys to probe the efficiency of
our estimator: (i) the ground-based LSST; (ii) the space-based SNAP;
(iii)a toy model, which could be a future cosmic 21-cm radiation
survey. Depending on the instrument used, four characteristics
describe a WL survey: the redshift distribution of the source
galaxies, $\mathcal{P}$(z), the angular concentration of source
galaxies, \nbar, the area of the sky covered, $\mathcal{A}$, and the
noise level of the intrinsic ellipticities of galaxies,
$\sigma_{\gamma}$. In all cases, we take $\sigma_{\gamma}$=0.3. For
observations of 21-cm emission at $z>5$, the lensing signal might be
extracted from shape or density correlations of discrete objects
(e.g. ``minihalos,'', \citet{2007RPPh...70..627B}) while at higher
redshifts the deflection field may be derived from an analysis of the
full 3D data cube in a manner analagous to the estimation of lensing
from the CMB itself (e.g. \citet{2006ApJ...653..922Z}). We retain the
notation of $\sigma_\gamma$ and $\bar n$ even in the absence of
discrete sources, since the quantity $\sigma^2_\gamma / \bar n$ still
has meaning as the noise level of the lensing reconstruction. The
attainable shape noise levels from 21-cm observations of $z>5$ are
highly dependent on the level of foreground contamination and the
history of reionization, so we will retain the simple $\bar n$
parameterization in this paper. For LSST and SNAP we employ the source
distribution introduced by \citet{1995ApJ...449L.105S}:
\be
\mathcal{P}(z)=\frac{1}{2 z_{0}^{3}}\,{\exp(-z/z_{0})}.
\ee
We normalize this distribution so
that $$\int_{0}^{\infty}dz\mathcal{P}(z)=\bar n\,.$$ The mean redshift of
this distribution is $\langle z \rangle=3z_{0}$, where $z_{0}$ is a
survey parameter. For LSST, we take $z_{0}=0.33$, \nbar=30 \gal and
$\mathcal{A}$=20000 deg$^{2}$. For SNAP, $z_{0}=0.5$, \nbar=100 \gal and
$\mathcal{A}$=1000 deg$^{2}$. For the toy model we adopt a "box"
(Heaviside) source galaxy distribution, i.e. all source bins have
equal number of source galaxies:
\be
\mathcal{P}(z)=\scriptstyle{\mathcal U}\ts{(z_{max}-z)}\,.
\ee
Since we are trying to find an optimal WL survey for our projected
potential estimator, we consider \zmax and \nbar as free parameters
for the toy model: \zmax ranges from 1 to 100 and \nbar from 4 to 250
\gal. However, we do fix $\mathcal{A}$=20000
deg$^{2}$, just like for LSST; this seems a reasonable choice, which
also allows us to compare the delensing efficiencies of the two
surveys.
\begin{figure*}
\centering
\begin{minipage}[t]{0.45\linewidth}
\includegraphics[scale=0.55]{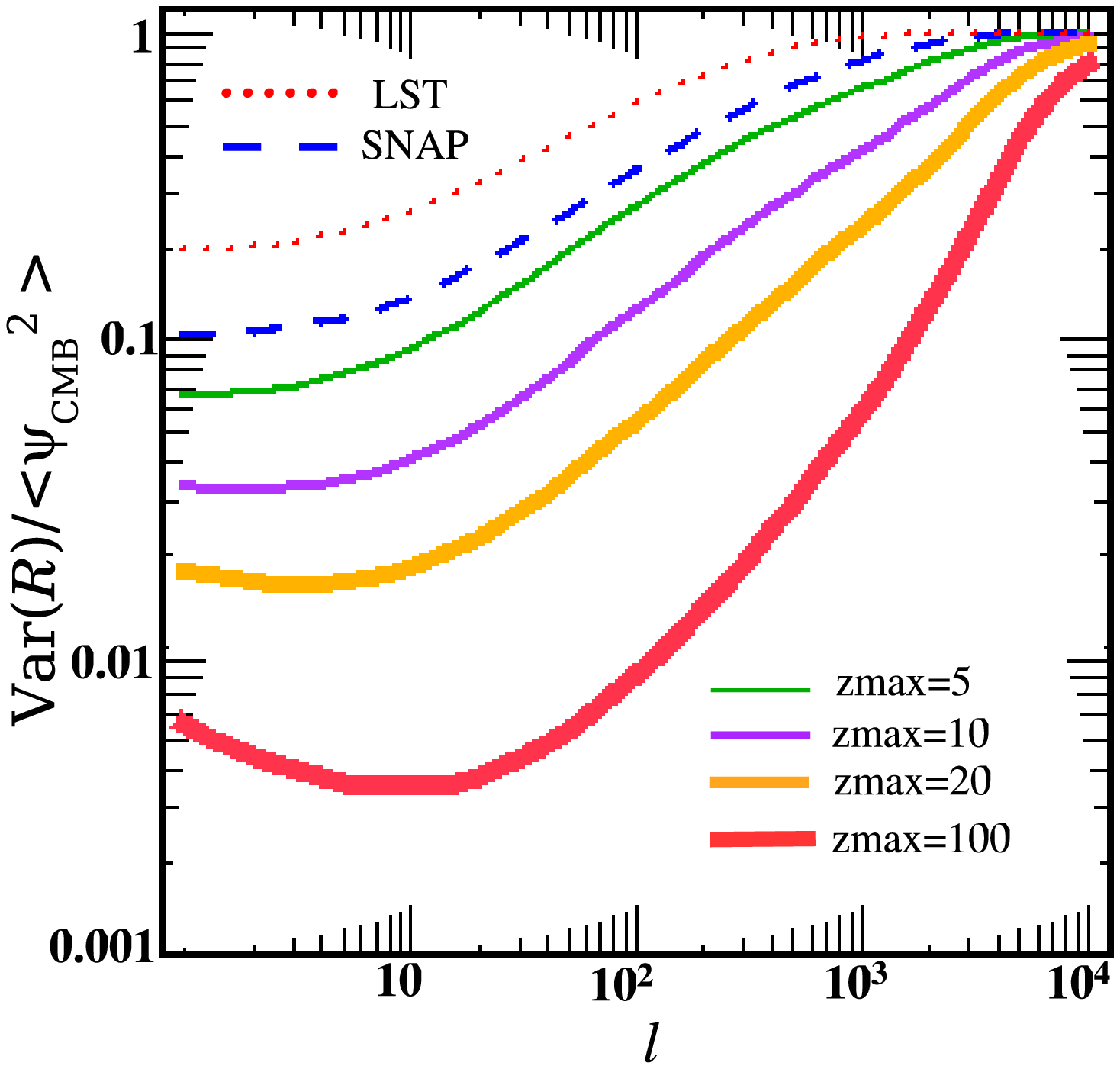}
\caption{The residual variance of the CMB lensing potential relative
  to its original value. LSST is the red dotted line, SNAP is the blue
  dashed line and 
  the solid lines are for the box distribution. For the latter case,
  \nbar=100 \gal and the thickness of the lines increases with
  the redshift depth of the survey.}
\label{fig:Res}
\end{minipage}%
\hspace{1.cm}
\begin{minipage}[t]{0.45\linewidth} 
\includegraphics[scale=0.56]{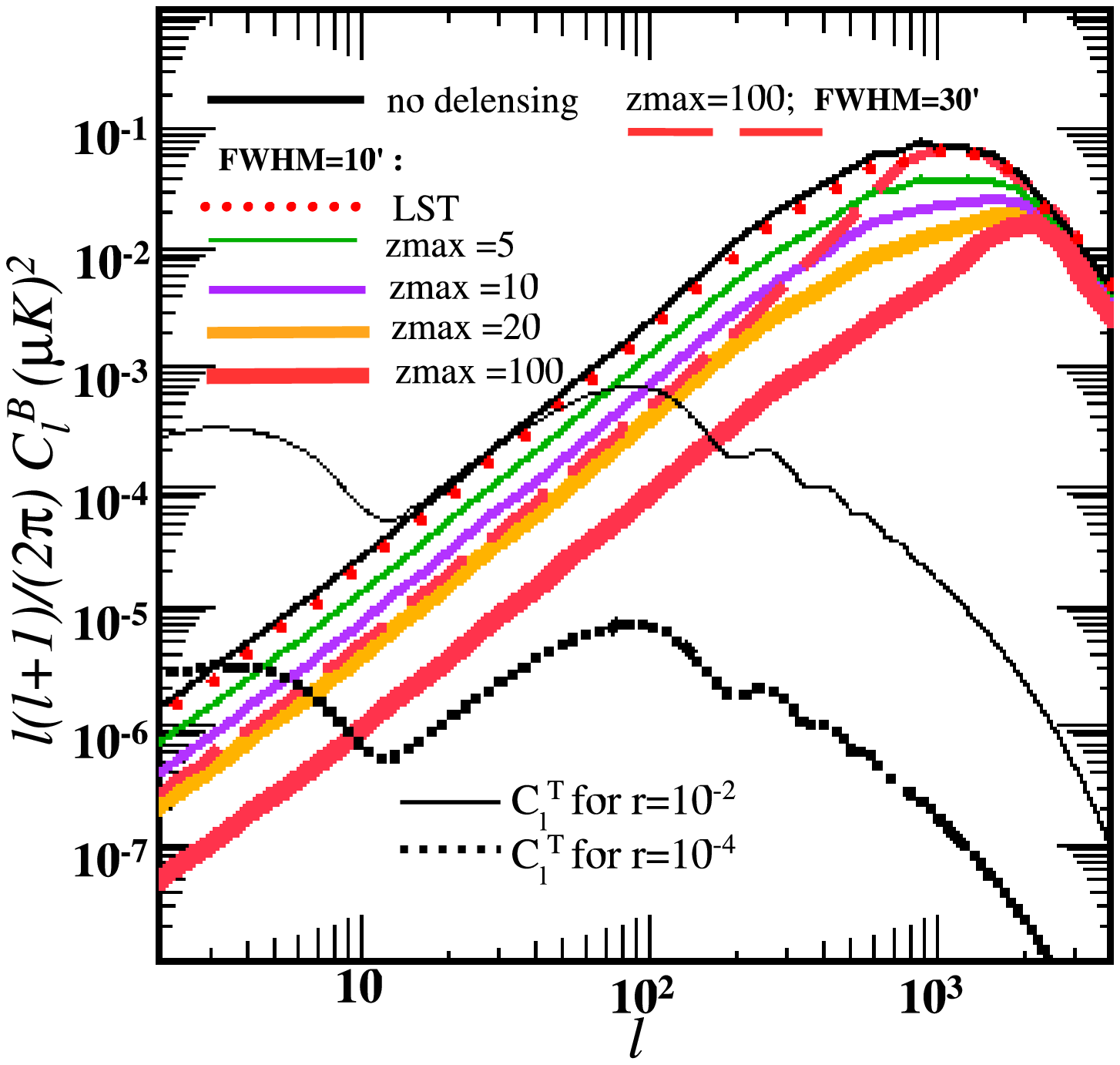}
\caption{The power spectra of the tensor $B$ mode, the WL $B$-mode
  contamination and its residual after delensing. The black dotted and
  thin solid lines are the tensor $B$ mode power spectra for
  $r=10^{-4}$ and $r=10^{-2}$, respectively. The upper solid black
  line is the full WL $B$ mode power and the rest of the lines are the
  residual contamination power spectra after delensing with our
  optimal estimator. The latter correspond to $w^{-1/2}=1 \mu$K arcmin and $\theta_{\rm\scriptscriptstyle{FWHM}}$=10 arcmin. The red, long-dashed curve corresponds to $\theta_{\rm\scriptscriptstyle{FWHM}}$=30 arcmin and \zmax=100.}
 \label{fig:Bmode}
\end{minipage}
\end{figure*}
\par We now specify a few noise definitions for
CMB polarization experiments, which will be relevant in section
\S\ref{IV}. We adopt the conventions of \citet{1995PhRvD..52.4307K}. We
assume that different pixels in the polarization map have
uncorrelated noise with uniform variance $\sigma^{2}_{pix}$. For one
detector, $\sigma^{2}_{pix}=\frac{\ts s^{2}}{\ts t_{pix}},$ where $s$ is the
sensitivity of the detector and $t_{pix}$ is the time spent on the
pixel. Then the noise power spectrum is:
\be
C_{l}^{N}\,\equiv \langle a_{lm}^{N}\,a_{l'm'}^{N*}\rangle=w^{-1}\,(W_{l}^{beam})^{-2}\,\delta_{ll'}\delta_{mm'}\,.
\label{eq:CmbNoise}
\ee
Here, $w$ is the weight per solid angle, given by:
$w^{-1}=\sigma^{2}_{pix}\,\Omega_{pix}$. The weight per solid angle
has the advantage that is independent of the beam size for fixed
survey time and survey area. We choose the beam to be Gaussian,
therefore the beam window function is:
$W^{beam}_{l}=\exp(-l^{2}\,\sigma_{beam}^{2}/2),$ where
$\sigma_{beam}^{2}=\ts{\theta_{fwhm}^{2}}/\ts{8\ln2}$.
\par We define a weighted estimator for the measured $E$ mode, with
weights chosen so that the variance of the residual of this estimator
is minimal. In this case the estimator $\hat E(\mx{\bm$l$})$ is just a scalar weight 
$\alpha_l^E$ times the noisy observation of $E(\mx{\bm$l$})$.
Following the same steps as in the case of the lensing
potential estimator, the weights and the variance of the
residual of the measured $E$ mode estimator are:
\ba
\alpha_{l}^{E} &=&\frac{C_{l}^{E}}{C_{l}^{E}+C_{l}^{N}}\,, \nonumber\\
\mx{\rm{Var}}\,\mathcal{R}^{E}(l)&=&\frac{C_{l}^{E}\,C_{l}^{N}}{C_{l}^{E}+C_{l}^{N}}\,.
\ea
\par All the results in
this paper correspond to a fiducial cosmology in accord with the third
year WMAP data release, i.e.  \citet{2007ApJS..170..377S}: flat
$\Lambda$CDM universe with $\Omega_{m}h^{2}$=0.128,
$\Omega_{b}h^{2}$=0.02, $h$=0.73, $n_{s}$=0.96, $\tau$=0.089 and
$\sigma_{8}$=0.76.
\par Figure~\ref{fig:Res} shows the variance of the residual
$\mathcal{R}(l)$, scaled by the CMB projected potential power
spectrum, for the three examples that we consider. The SNAP (LSST)
estimator removes up to 90\% (80\%) of the CMB projected power for
the lowest multipoles (l\st100) and only up to 20\% (10\%) for
l\st1000. This happens largely for two reasons. First and most
important, LSST and SNAP do not have source galaxies at redshifts
higher than \ax 4 and \ax 6, respectively. Therefore the potential
fluctuations between the upper-limit redshifts 4(6) and 1100 cannot
be reconstructed; since the horizon is much smaller at
these redshifts than today, the small scale power suffers greatly from
this effect. Second, the shot noise starts to dominate the lensing
signal at l\ax1000, so the reconstruction is bound to fail at high
multipoles, independent of the cosmological information provided by
the source bins. It is easier to separate the impact of shot noise on
the estimator from that of the redshift depth of the WL survey if we
look at our third fiducial survey, the box distribution. The green,
solid line shows the performance of the optimal estimator using the
box distribution with \zmax=5 and \nbar=100 \gal. The total shot noise
is the same as for SNAP, but the reconstruction is improved because
there are more high-redshift sources. If we take \zmax=20 and keep the
same total angular concentration of galaxies, then the estimator
improves by a factor of \ax 3 at low multipoles. We expect that by
taking \zmax$\rightarrow z_{\scriptscriptstyle \rm{CMB}}$ and
\nbar$\rightarrow\infty$, we can recover entirely the projected
potential to the CMB. The red solid curve in Fig.~\ref{fig:Res},
correponding to a redshift depth of 100 for the box distribution,
seems to confirm this expectation. The estimator's efficiency
increases relatively smoothly with the value of \nbar: the difference
between having 100 and 250 \gal leads to at most a factor of 2 in
the estimator's efficiency for \zmax=100. This difference becomes
smaller with lower \zmax. We conclude that for the purpose of the CMB
projected potential reconstruction, going to high redshift in a WL
survey is much more important than trying to resolve a high number
density of source galaxies. In section \S\ref{IV} we give a more
quantitative flavor to this conclusion and discuss what values of
\nbar and \zmax we need in a WL survey to significantly delens a
$B$-mode map.
\section{Detecting the tensor-to-scalar ratio}
\label{IV}
\subsection{The Null-Hypothesis Test}
\label{IVa}
\par We would like to answer the following question: given a CMB
survey, how much can we improve the detectability of the
tensor-to-scalar ratio $r$ if we use our projected-potential estimator
to delens the CMB? There are several factors that influence the answer
to this question: the sensitivity and angular resolution of the CMB
experiment that measures the B modes, the noise level and redshift
depth of the WL survey that provides the information for our potential
estimator and, finally, the true value of $r$ itself. The statistical
significance of detection for some $r>0$ is estimated from the
$\chi^{2}$ statistic that would be obtained under the null hypothesis
$r=0$.  The underlying approximation is that the delensed B mode obeys
Gaussian statistics, which would be true for perfect delensing and no
foregrounds. We choose the scenario $r=0$ to be our fiducial model. We
compare other models, with $r\neq 0$ to the fiducial one and we
determine the minimum $r$ statistically distinguishable from $r=0$.
The delensed $B^{del}(\mx{\bm$l$})$ will be constructed by subtracting
from the lensed $\tilde B(\mx{\bm$l$})$ an estimate of the two lensing
terms in equation~(\ref{eq:lensedB}).  Accounting for the fact that we
do not have access to the real lensing potential and $E$ mode, but
their estimators, the delensed field is:
\ba
\lefteqn{B^{del}(\mx{\bm$l$})=B(\mx{\bm$l$})\!-\!\!\int\frac{d^{2}l'}{2\pi}f\left(\mx{\bm$l'$,\,\bm$l$}\right)\left[\psi(\mx{\bm$L$})E(\mx{\bm$l'$})-\hat{\psi}(\mx{\bm$L$})\hat{E}(\mx{\bm$l'$})\right]}\hspace{0.3in}
\nn
& & -\!\frac{1}{2}\!\!\int\!\!\frac{d^{2}l'}{2\pi}\!\frac{d^{2}l''}{2\pi}g\left(\mx{\bm$l'$,\,\bm$l''$,\,\bm$l$}\right)[\,\psi(\mx{\bm$l''$})\psi^{*}(\mx{\bm$l''$}\!\!-\mx{\bm$L$})E(\mx{\bm$l'$})\nn
& & -\hat{\psi}(\mx{\bm$l''$})\hat{\psi}^{*}(\mx{\bm$l''$}\!\!-\mx{\bm$L$})\hat{E}(\mx{\bm$l'$})\,]+B^{N}(\mx{\bm$l$}).
\label{eq:delBmode}
\ea
We include the
measurement noise $B^{N}$, whose variance is given by
equation~(\ref{eq:CmbNoise}).
The variance of the measured, delensed $B$ mode follows from the above:
\be
C_{l}^{B,\,del}=rC_{l}^{tensor}+C_{l}^{N}+C_{l}^{\mx{$\scriptscriptstyle{WL}$},\,del}.
\label{eq:delBpower1}
\ee
We can express the residual power left from delensing,
$C_{l}^{\mx{$\scriptscriptstyle{WL}$},\,del}$, in terms of the variance of the
residuals of $\hat E$ and $\hat \psi$, and the power spectra of the
real lensing potential and $E$ mode:
\ba
\lefteqn{C_{l}^{\mx{$\scriptscriptstyle{WL}$},\,del}=\int\frac{d^{2}l'}{2\pi}\left|f\left(\mx{\bm$l'$,\,\bm$l$}\right)\right|^{2}\,[\,C_{L}^{\psi}\,\mx{\rm{Var}}\mathcal{R}_{l'}^{E}\,+
\mx{\rm{Var}}\mathcal{R}_{L}^{\psi}\,C_{l'}^{E}-} \nn
& & -\mx{\rm{Var}}\mathcal{R}_{L}^{\psi}\,\mx{\rm{Var}}\mathcal{R}_{l'}^{E}\,]+\frac{1}{4}\int\!\!\frac{d^{2}l'}{2\pi}\!\frac{d^{2}l''}{2\pi}|g\left(\mx{\bm$l'$,\,\bm$l''$,\,\bm$l$}\right)|^{2}\,\times \nn
& &\{C_{l''}^{\psi}\,C_{|\mx{\bm{$\scriptstyle l''$}-\bm{$\scriptstyle L$}}|}\,\mx{\rm{Var}}\mathcal{R}_{l'}^{E}+(\,C_{l'}^{E}-\mx{\rm{Var}}\mathcal{R}_{l'}^{E}\,)\,[\,C_{l''}^{\psi}\,\mx{\rm{Var}}\mathcal{R}_{|\mx{\bm{$\scriptstyle l''$}-\bm{$\scriptstyle L$}}|}^{\psi}+ \nn
& & +C_{|\mx{\bm{$\scriptstyle l''$}-\bm{$\scriptstyle L$}}|}\,\mx{\rm{Var}}\mathcal{R}_{l''}^{\psi} 
-\mx{\rm{Var}}\,\mathcal{R}_{l''}^{\psi}\,\mx{\rm{Var}}\mathcal{R}_{|\mx{\bm{$\scriptstyle l''$}-\bm{$\scriptstyle L$}}|}^{\psi}\,]\,\}.
\label{eq:delBpower2}
\ea
The last integral in the above equation, containing terms of the
second order in the potential power, yields a much smaller
contribution to $C_{l}^{\mx{$\scriptscriptstyle{WL}$},\,del}$ compared to the first
term. For computational simplicity, we choose therefore to neglect
this last term; all the results following have been obtained in this
approximation.
Let us write the $\Delta\chi^{2}$ for the measured
and delensed B-mode power spectrum:
\be
\Delta\chi^{2}(r)\!=\!\!\sum_{l}\!\!\left[C^{B,\,del}_{l}(r)\!-\!C^{B,\,del}_{l}(r=0)\right]^{\!2}\!\!\!\mx{\rm{Var}}\!\left[C^{B,\,del}_{l}(r)\right]^{-1}_{r=0}.
\label{eq:deltachi}
\ee
 The variance of the delensed B-mode power spectrum is:
\be
\mx{\rm{Var}}(C^{B,\,del}_{l}(r))=\frac{2}{(2l+1)f_{sky}}\left[r C_{l}^{tensor}+C_l^{N}+C^{\mx{$\scriptscriptstyle{WL}$},\,del}_{l}\right]^{2}.
\ee
Here $f_{sky}$ denotes the fraction of sky covered by the CMB
mission. To be able to use our lensing potential estimator, the CMB
experiment must survey the same area of the sky as the WL experiment.
Using the above equation in the case of the fiducial model $r=0$, we
rewrite Eq.~(\ref{eq:deltachi}) as:
\ba
\Delta\chi^{2}(r)=\sum_{l}\frac{2l+1}{2}f_{sky}\left[\frac{r\,C_{l}^{tensor}}{C^{WL,del}_{l}+C_l^{N}}\right]^{2}.
\label{eq:deltachi2}
\ea
In order to find the smallest $r$ discernible from the fiducial value
of 0, we need to solve the following equation:
\be
\Delta\chi^{2}(r)=\alpha^{2} ,
\label{eq:rmin} 
\ee
where $\alpha$ sets the confidence level. All our results are computed
for $\alpha$=2, which corresponds to a 95\% confidence level for a
single free parameter $r$.
\begin{figure*}
\centering
\begin{minipage}[t]{0.45\linewidth}
\includegraphics[scale=0.55]{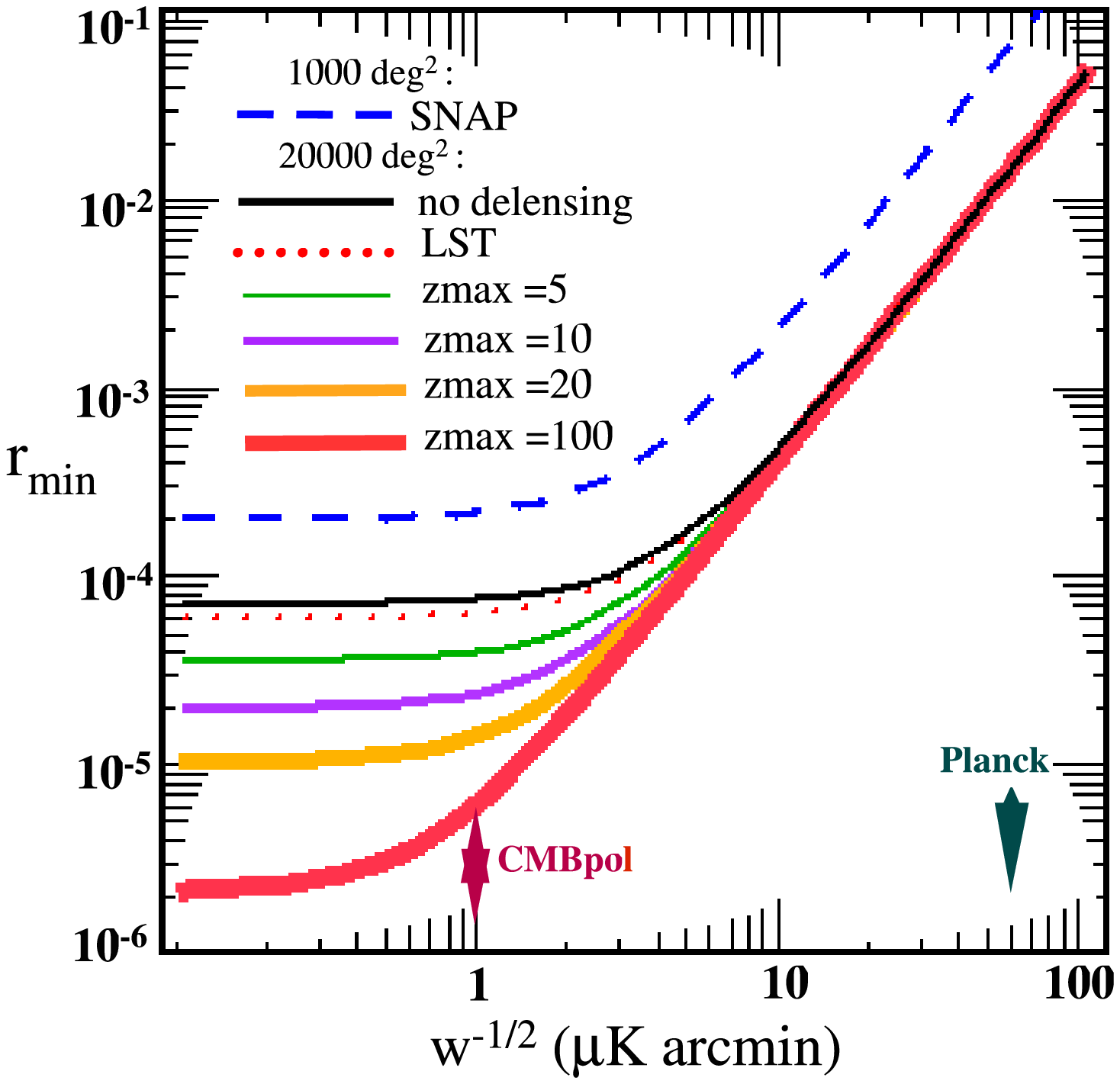}
\caption{The minimum tensor-to-scalar r distinguishable from 0 at 95\%
  confidence level as a function of CMB detector noise. All curves are
  for \nbar=100 \gal, with the exception of the LSST line, for which
  \nbar=30 \gal. The upper blue dashed curve is for SNAP delensing
  ($\mathcal{A}$=1000 $\rm{deg^{2}}$). The rest of the lines
  correspond to a 20000 $\rm{deg^{2}}$ sky coverage and match the
  delensing cases presented in Fig.~\ref{fig:Bmode}. We have also
  indicated the approximate polarization noise levels for the Planck
  and CMBPol missions.}
\label{fig:rmin}
\end{minipage}%
\hspace{1.cm}
\begin{minipage}[t]{0.45\linewidth} 
\includegraphics[scale=0.56]{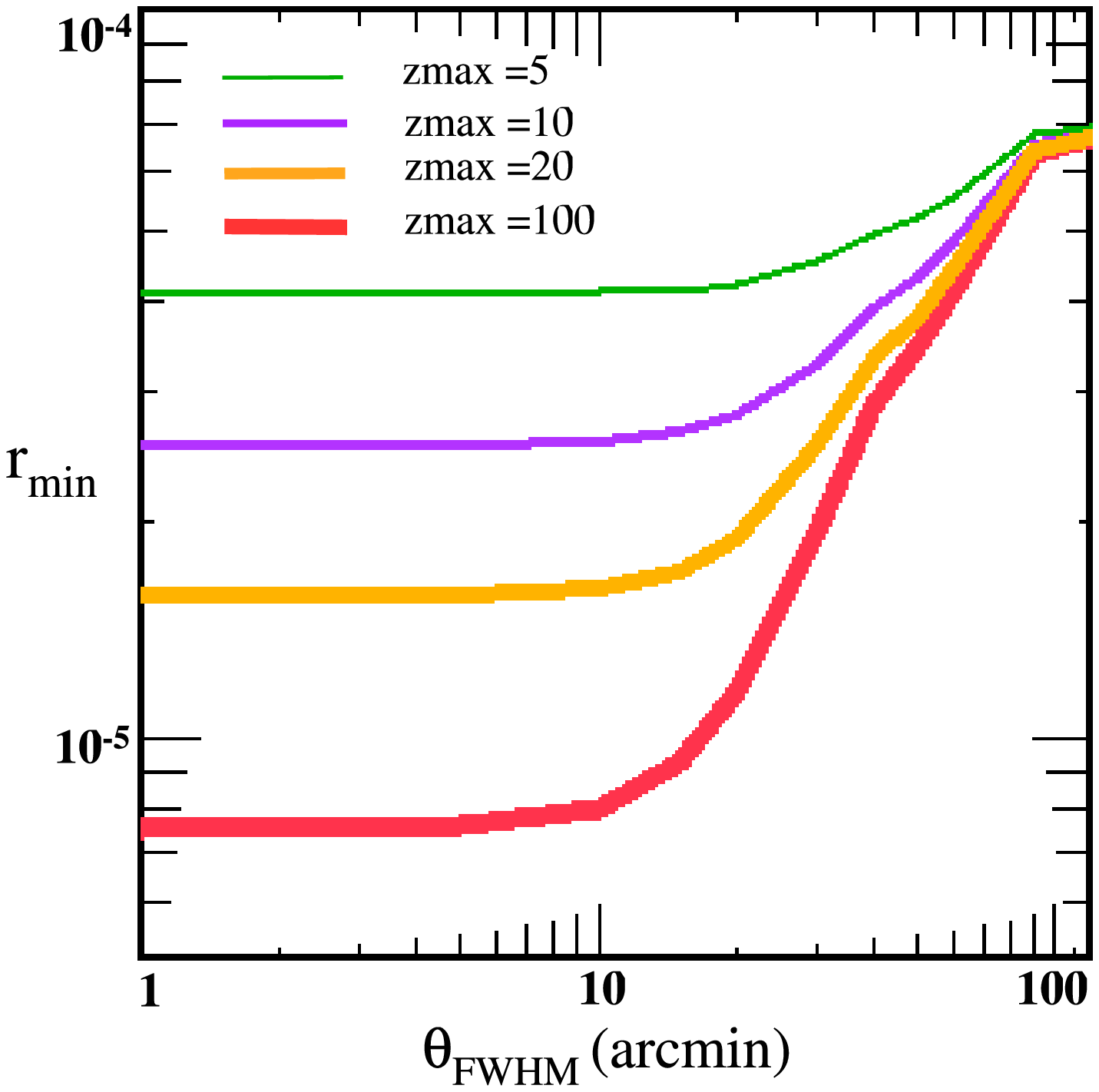}
\caption{Dependence of $r_{min}$ on the beam size. The curves are for
  a box distribution with \zmax=5, 10, 20 and 100, \nbar=100 \gal and
  the noise level of CMBPol.}
 \label{fig:beams}
\end{minipage}
\end{figure*}
\subsection{Results}
\label{IVb}
Figure~\ref{fig:Bmode} shows the lensing $B$ mode, given by
Eq.~\ref{eq:lensedClB2} and the tensor $B$ mode for two values of
$r=10^{-2}, 10^{-4}$. The latter was computed using the publicly
available code CAMB, i.e. \footnote{\tt{http://camb.info/}},
\citet{2000ApJ...538..473L}. Also shown is the residual lensing power,
given by Eq.~\ref{eq:delBpower2}, when delensing using LSST and box
estimators has been applied. With one exception, the plots correspond
to a beam $\theta_{\rm\scriptscriptstyle{FWHM}}$=10 arcmin; for all of them, the
noise level is $w^{-1/2}=1 \mu$K arcmin, i.e the expected CMBPol noise
level.  \par We can see that LSST delensing can reduce the lensing
$B$-mode power by a factor of \ax 1.3 (\ax 2 for SNAP, although not
shown in the figure), while a box with \zmax=20 and \nbar=100 \gal
leaves a residual power \ax 8 times lower than the original
contamination. Although such high redshifts are not attainable by a WL
galaxy survey, we would like to illustrate the potential benefits of a
21-cm emission survey. There is further improvement as we go to higher
$z_{max}$: for \zmax=100 and the same value of \nbar, the residual
lensing power is \ax 37 times lower than the full lensing $B$-mode
power. Increasing the beam width has a dramatic effect: the
long-dashed red line is for the same box with \zmax=100, but a beam
with $\theta_{\rm\scriptscriptstyle{FWHM}}$=30 arcmin; compared to the 10 arcmin
beam, the delensing efficiency has dropped by a factor of 6.  This is
understood as follows: the lensed $B$-mode at $l\sim$10--100 arises in
part from the beating of $E$-mode power at $l\sim1000$ with deflection
power at similar $l$. The delensing must know both these $E$ and
deflection fields at $l\approx1000$ in order to succeed, hence the CMB
experiment must resolve these modes.
\par Naturally, we would like to probe now how much delensing with the
above-mentioned WL survey models impacts the detection of tensor
modes. In figure~\ref{fig:rmin} we plot the minimum tensor-to-scalar
ratio distinguishable from 0 at 95\% confidence level as a function of
the CMB experimental noise for a fixed beam
$\theta_{\rm\scriptscriptstyle{FWHM}}$=10 arcmin. $r_{min}$ is obtained by solving
Eq.~\ref{eq:rmin} for some of the delensing scenarios discussed
earlier and the noise is in the form of weight per solid angle,
$w^{-1/2}$. The upper thin, solid, black curve corresponds to the case
when no delensing is done. We also show (turquoise and red arrows) the
approximate noise levels $w^{-1/2}=60$ and 1~$\mu$K~arcmin for two
future missions: the Planck satellite
\footnote{\tt{http://www.rssd.esa.int/Planck}} and CMBPol. The noise
level for the latter was estimated by assuming 1000 detectors with a
sensitivity of 50 $\mu$K$\sqrt{\rm{sec}}$, sky coverage of 20000
$\rm{deg^{2}}$ and 5 years of operation. 
\begin{figure}
\includegraphics[scale=0.56]{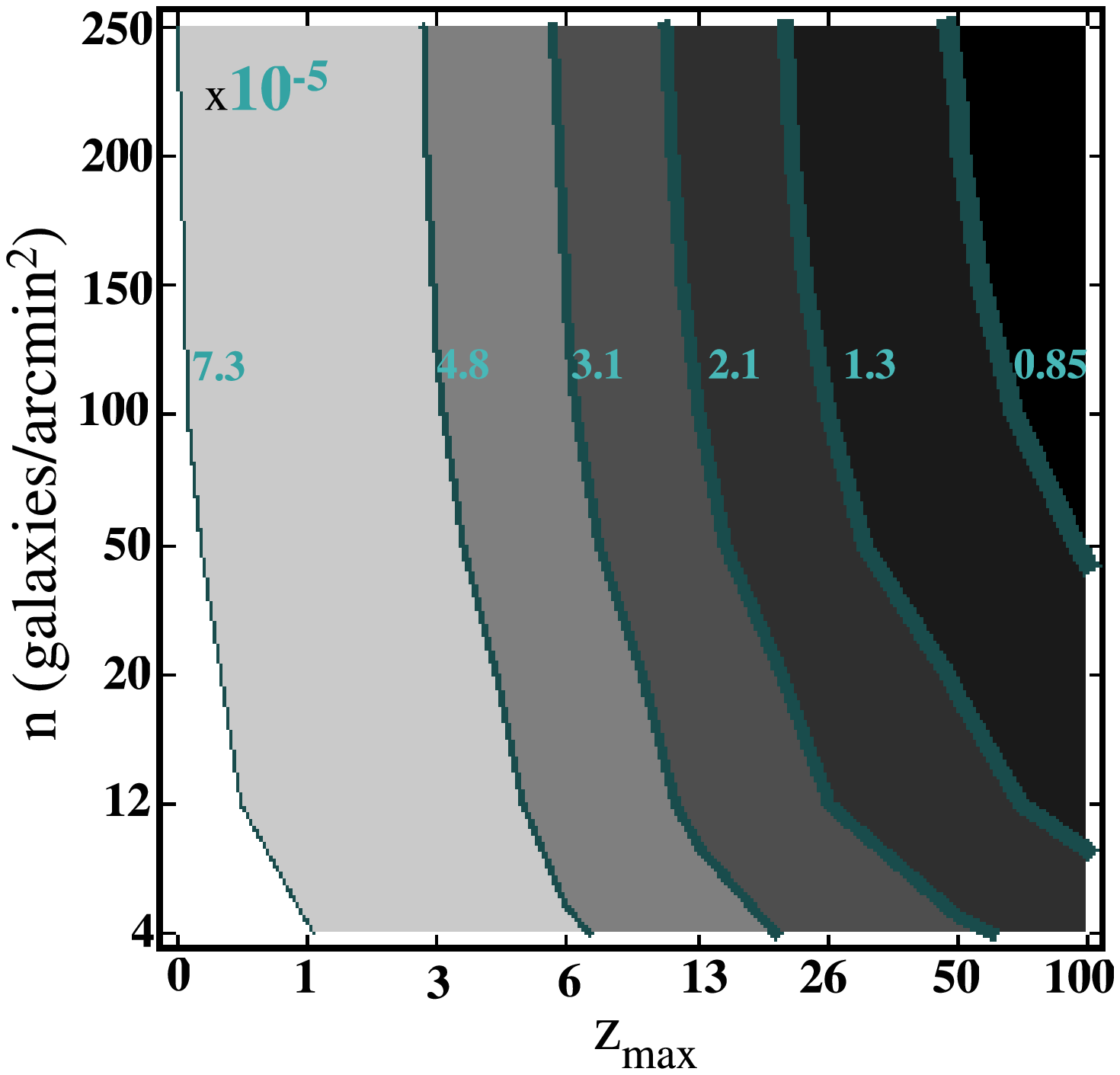}
\caption{$r_{min}$ as a function of the WL survey parameters, \zmax
  and $\bar{n}$. The contours are for a box distribution, the noise
  level of CMBPol and $\theta_{\rm\scriptscriptstyle{FWHM}}$=10 arcmin. Their
  increasing thickness depicts the change in the detectable tensor $B$
  mode amplitude as we transit from a weak delensing case (\zmax=1,
  \nbar=250 \gal) to very good delensing (\zmax=100, \nbar=250 \gal).}
 \label{fig:contours}
\end{figure}
Delensing is relevant for the detection of tensor $B$ modes only if
the instrumental noise is low enough, $w^{-1/2}<10 \mu$K arcmin. And
this is an optimistic value, since our analysis does not take
foregrounds into account. While CMBPol is meant to reach such low
noise levels, delensing will not, most probably, be a concern for
Planck. This is in accord with the findings of the Planck Science
Team.  For low noise levels, when delensing does play a part,
figure~\ref{fig:rmin} reinforces our earlier conclusions: the
difference between the case of no delensing and that of a box
distribution going to \zmax=20 translates approximately into a factor
of 7 difference in the corresponding $r_{min}$ for the lowest detector
noise that we consider, $w^{-1/2}=0.1\mu$K arcmin. \zmax=100 gains
another order of magnitude in the detection threshold of the tensor
$B$ modes, for the same weight per solid angle.  This is quite
idealized, however: for the technically ambitious CMBPol, at
$w^{-1/2}=1\mu$K arcmin, the extension to \zmax=100 gains only a
factor of 3 over \zmax=20.  The 21-cm signal does not exist
significantly above $z\approx 100$ since the hydrogen spin temperature
is expected to match the CMB temperature at earlier epochs, for instance see \citet{2007RPPh...70..627B} and \citet{2006ApJ...653..922Z}.

We note that $r_{min}$ corresponding to LSST delensing is lower by a
factor of \ax 3 than the SNAP $r_{min}$, despite the better
performance of the SNAP lensing potential estimator. This is due to
LSST's superior sky coverage and the presumed larger CMB sky
coverage. We emphasize that the improvement in $r_{min}$ for LSST and
SNAP delensing relative to the case of no delensing (corresponding to
their respective survey areas of 20000 $\rm{deg^{2}}$ and 1000
$\rm{deg^{2}}$) is of only \ax 20\% and 50\%, respectively. We
expected this result: the potential modes with angular scales of $l$
\ax1000 mix with $E$ power and generate the low-$l$ $B$ mode polarization
power; but such high-$l$ lensing potential modes are poorly
reconstructed by both LSST and SNAP.  \par Figure~\ref{fig:beams}
illustrates the importance of the beam choice for our delensing
method. For the CMBPol noise level $w^{-1/2}=1 \mu$K arcmin, we plot
$r_{min}$ as a function of beam size, for the same box models
mentioned in Fig.~\ref{fig:rmin}. If $\theta_{\rm\scriptscriptstyle{FWHM}}$ \st 10
arcmin, $r_{min}$ stays constant for all models considered. But once
$\theta_{\rm\scriptscriptstyle{FWHM}}>$10 arcmin, the delensing efficiency drops
steeply and for $\theta_{\rm\scriptscriptstyle{FWHM}}=2^{\circ}$, there is nothing
to be gained from delensing.  \par We now explore the importance of
the WL free parameters \zmax$\,$ and \nbar$\:$ for the tensor $B$
modes detection. Figure~\ref{fig:contours} is a contour plot of
$r_{min}$ as a function of redshift depth and angular concentration of
source galaxies. All contours are plotted using the box distribution
for a CMBPol-level of detector noise, $\theta_{\rm\scriptscriptstyle{FWHM}}$=10
arcmin and a 20000 $\rm{deg^{2}}$ sky coverage. This plot confirms our
findings from section \S\ref{IIIc}. Once a certain source density is
reached (\nbar\ax 50 \gal at \zmax\st10), there is no significant
improvement in $r_{min}$ if we continue to increase \nbar. The
redshift depth is the dominant parameter of the WL survey: even with a
concentration as low as 10 \gal, if we reach sources at z=15,
$r_{min}$ has a factor of 2 improvement over the case of no
delensing. Only if we reach \zmax\gt 100 can we significantly
improve $r_{min}$ by increasing \nbar above 50 \gal.
\subsection{Related Work in the Literature}
\label{IVc}
\par\citet{2002PhRvL..89a1303K} investigate the detectability of
inflationary gravitational waves using CMB observables to delens. They
apply the quadratic estimator of \citet{2002ApJ...574..566H} to a CMB
polarization experiment and thus reconstruct the lensing
potential. They find that for a noise-free CMB map, the lensing
$B$-mode power can be reduced by an order of magnitude, which
corresponds to a similar reduction in $r_{min}$. 
\par \citet{2002PhRvL..89a1304K} apply the same quadratic estimator to
temperature maps and find similar results as
\citet{2002PhRvL..89a1303K}.
\par\citet{2005PhRvL..95u1303S} reconstruct the lensing potential using
the above-mentioned quadratic estimator applied to 21-cm temperature
fluctuations. They account for two distinct sources of noise that
affect their estimator: the reconstruction noise, and the 
unreconstructed lensing due to mass between the 21-cm source redshift and
the CMB.
In the limit where the reconstruction noise
dominates, they find that a 21-cm survey, combined with a very
sensitive CMB survey, could reduce the detection threshold of the
tensor $B$-mode power by an order of magnitude, compared to the no
delensing case. The authors also consider the limit where the partial
delensing bias dominates. In this limit, the reduction in the lensing
contamination power is the same as what we find if we take
\nbar$\rightarrow\infty$.
\par\citet{2004PhRvD..69d3005S} show that the reconstruction of the
lensing potential using the maximum likelihood estimator applied to
polarization maps can reduce the power of the lensing $B$ mode by at
least a factor of 40. They conclude that delensing is not a
fundamental limit to the detection of inflationary gravitational
waves, if high-resolution CMB maps cleaned of foreground emission are
available. 
\section{Conclusions}
\label{V}
In this paper we have examined the possibility of delensing $B$-mode
polarization maps using galaxy WL surveys. We have proposed a weighted
combination of projected potential estimators to different source
redshift bins which optimally reconstructs the projected potential
seen by the CMB. We have used three fiducial surveys to exemplify our
estimator: LSST, SNAP and a generic survey relevant mostly to future
21-cm studies. These examples have different source redshift
distribution, source density \nbar, and sky coverage $f_{\rm sky}$,
and have enabled us to test the effect of each of these factors on the
lensing potential estimator and also on its ability to reduce the WL
contamination of $B$ mode maps. Using a $\Delta \chi^{2}$ test for the
delensed $B$ mode field, we have determined the minimum value of the
tensor-to-scalar ratio statistically distinguishable from 0 in
optimally delensed data. Throughout this paper we have ignored the
polarized foreground contamination and we have also made the
assumption of Gaussianity for the delensed $B$ mode map. While this
assumption may be inaccurate, especially for the cases when the
efficiency of delensing is low, we expect the qualitative results of
our work to hold.  \par The lensing potential estimator is sensitive
mostly to the redshift depth of the WL survey and reconstructs best
the large angular scale multipoles. The performance of the estimator
improves continually as higher redshift source galaxies are used. In
the limit \zmax$\rightarrow z_{\scriptscriptstyle\rm CMB}$ and
\nbar$\rightarrow\infty$, the reconstruction is perfect. An experiment
like SNAP recovers \ax 90\% of the CMB projected potential power for
$l\leq 100$ and \ax 20\% for $l\leq 1000$. The performance of LSST is
a little worse, because it detects sources at an average redshift of
$\langle z \rangle$=1, compared to $\langle z \rangle$=1.5 of SNAP. A
box distribution (constant redshift distribution of source galaxies),
provides better CMB potential estimator: if we go as far as \zmax=20,
there is an order of magnitude improvement over SNAP, for the same
angular concentration of galaxies.

However, in this last case, the reduction in $r_{min}$, compared to
the case where no delensing is done, is only of a factor of \ax 7 for
$w^{-1/2}=0.1 \mu$K arcmin and $\theta_{\rm\scriptscriptstyle{FWHM}}$=10
arcmin. This happens because low-$l$ lensing $B$-modes are generated
by beating of gravitational potential modes and $E$-modes on scales
coresponding to $l$\ax1000, where the potential estimator is less
faithful.  In the case of LSST and SNAP, the reduction in $r_{min}$
relative to the no delensing case is only by \ax15\% and \ax50\% for
the same CMB noise level and beam.

We stress that delensing with foreground weak lensing offers
significant gains in $r_{min}$ (factors of 5 or more) only when three
conditions are met:
\begin{enumerate}
\item the noise in the CMB map is sufficiently low, $\sim 1\,\mu$K arcmin;
\item the beam size of the CMB map is $<20$ arcmin;
\item the lensing source distribution extends to $z\sim 20$ or greater.
\end{enumerate}
Delensing is not relevant to tensor mode detection if $w^{-1/2}>10
\mu$K arcmin or if $\theta_{\rm\scriptscriptstyle{FWHM}}$ \gt$2^{\,\circ}$. Also, there
is no advantage in lowering the beam size beyond
$\theta_{\rm\scriptscriptstyle{FWHM}}$=10 arcmin: for our delensing method, a 1
arcmin beam will do just as well as a 10 arcmin beam. This is perhaps
the most important feature of our delensing technique, as both the
quadratic estimator of \citet{2001ApJ...557L..79H} and the maximum
likelihood estimator proposed by \citet{2003PhRvD..68h3002H} require
beam sizes of 2-3 arcmin to yield their best performance. For the CMBPol
mission, (which might reach a noise of 1 $\mu$K arcmin) delensing with
a box of \zmax=20 results in $r_{min}\approx 2\times 10^{-5}$. If no
delensing is applied, then $r_{min}\approx 8\times10^{-5}$, a factor
of 4 worse. CMBPol detector noise keeps $r_{min}>7\times 10^{-6}$ even
with perfect delensing, so there is less incentive to acquire
delensing source planes above \zmax=20. Also for the CMBPol noise
level, we have investigated the impact of $\bar n$ and have concluded
that as long as we go to high redshift ($z_{max}>10$), even with a low
density of source galaxies, we can still improve $r_{min}$ by a factor
of a few.

\section*{Acknowledgments}
LM is supported by grant BEFS-04-0014-0018 from NASA. GMB acknowledges
additional support from Department of Energy grant
DOE-DE-FG02-95ER40893 and National Science Foundation grant
AST-0607667. This work has substantially benefitted from the suggestions of the
anonymous referee: we would like to thank him or her. We are very
grateful to Robert E. Smith and Jacek Guzik for their participation in
the CMB lensing review sessions. We thank Carlos Hernandez-Monteagudo,
Jeff Klein, Dan Swetz and Chris Semisch for useful discussions.


\begin{thebibliography}{25}
\expandafter\ifx\csname natexlab\endcsname\relax\def\natexlab#1{#1}\fi
\expandafter\ifx\csname bibnamefont\endcsname\relax
  \def\bibnamefont#1{#1}\fi
\expandafter\ifx\csname bibfnamefont\endcsname\relax
  \def\bibfnamefont#1{#1}\fi
\expandafter\ifx\csname citenamefont\endcsname\relax
  \def\citenamefont#1{#1}\fi
\expandafter\ifx\csname url\endcsname\relax
  \def\url#1{\texttt{#1}}\fi
\expandafter\ifx\csname urlprefix\endcsname\relax\def\urlprefix{URL }\fi
\providecommand{\bibinfo}[2]{#2}
\providecommand{\eprint}[2][]{\url{#2}}

\bibitem[{\citenamefont{{Zaldarriaga} and
  {Seljak}}(1999)}]{1999PhRvD..59l3507Z}
\bibinfo{author}{\bibfnamefont{M.}~\bibnamefont{{Zaldarriaga}}}
  \bibnamefont{and} \bibinfo{author}{\bibfnamefont{U.}~\bibnamefont{{Seljak}}},
  \bibinfo{journal}{\prd} \textbf{\bibinfo{volume}{59}},
  \bibinfo{pages}{123507} (\bibinfo{year}{1999}), \eprint{astro-ph/9810257}.

\bibitem[{\citenamefont{{Bernardeau}}(1998)}]{1998A&A...338..767B}
\bibinfo{author}{\bibfnamefont{F.}~\bibnamefont{{Bernardeau}}},
  \bibinfo{journal}{\aap} \textbf{\bibinfo{volume}{338}}, \bibinfo{pages}{767}
  (\bibinfo{year}{1998}), \eprint{astro-ph/9802243}.

\bibitem[{\citenamefont{{Hu}}(2001{\natexlab{a}})}]{2001PhRvD..64h3005H}
\bibinfo{author}{\bibfnamefont{W.}~\bibnamefont{{Hu}}}, \bibinfo{journal}{\prd}
  \textbf{\bibinfo{volume}{64}}, \bibinfo{pages}{083005}
  (\bibinfo{year}{2001}{\natexlab{a}}), \eprint{astro-ph/0105117}.

\bibitem[{\citenamefont{{Hu}}(2001{\natexlab{b}})}]{2001ApJ...557L..79H}
\bibinfo{author}{\bibfnamefont{W.}~\bibnamefont{{Hu}}},
  \bibinfo{journal}{\apjl} \textbf{\bibinfo{volume}{557}}, \bibinfo{pages}{L79}
  (\bibinfo{year}{2001}{\natexlab{b}}), \eprint{astro-ph/0105424}.

\bibitem[{\citenamefont{{Guzik} et~al.}(2000)\citenamefont{{Guzik}, {Seljak},
  and {Zaldarriaga}}}]{2000PhRvD..62d3517G}
\bibinfo{author}{\bibfnamefont{J.}~\bibnamefont{{Guzik}}},
  \bibinfo{author}{\bibfnamefont{U.}~\bibnamefont{{Seljak}}}, \bibnamefont{and}
  \bibinfo{author}{\bibfnamefont{M.}~\bibnamefont{{Zaldarriaga}}},
  \bibinfo{journal}{\prd} \textbf{\bibinfo{volume}{62}},
  \bibinfo{pages}{043517} (\bibinfo{year}{2000}), \eprint{astro-ph/9912505}.

\bibitem[{\citenamefont{{Hu} and {Okamoto}}(2002)}]{2002ApJ...574..566H}
\bibinfo{author}{\bibfnamefont{W.}~\bibnamefont{{Hu}}} \bibnamefont{and}
  \bibinfo{author}{\bibfnamefont{T.}~\bibnamefont{{Okamoto}}},
  \bibinfo{journal}{\apj} \textbf{\bibinfo{volume}{574}}, \bibinfo{pages}{566}
  (\bibinfo{year}{2002}), \eprint{astro-ph/0111606}.

\bibitem[{\citenamefont{{Kesden} et~al.}(2003)\citenamefont{{Kesden}, {Cooray},
  and {Kamionkowski}}}]{2003PhRvD..67l3507K}
\bibinfo{author}{\bibfnamefont{M.}~\bibnamefont{{Kesden}}},
  \bibinfo{author}{\bibfnamefont{A.}~\bibnamefont{{Cooray}}}, \bibnamefont{and}
  \bibinfo{author}{\bibfnamefont{M.}~\bibnamefont{{Kamionkowski}}},
  \bibinfo{journal}{\prd} \textbf{\bibinfo{volume}{67}},
  \bibinfo{pages}{123507} (\bibinfo{year}{2003}), \eprint{astro-ph/0302536}.

\bibitem[{\citenamefont{{Kesden} et~al.}(2002)\citenamefont{{Kesden}, {Cooray},
  and {Kamionkowski}}}]{2002PhRvL..89a1304K}
\bibinfo{author}{\bibfnamefont{M.}~\bibnamefont{{Kesden}}},
  \bibinfo{author}{\bibfnamefont{A.}~\bibnamefont{{Cooray}}}, \bibnamefont{and}
  \bibinfo{author}{\bibfnamefont{M.}~\bibnamefont{{Kamionkowski}}},
  \bibinfo{journal}{Physical Review Letters} \textbf{\bibinfo{volume}{89}},
  \bibinfo{pages}{011304} (\bibinfo{year}{2002}), \eprint{astro-ph/0202434}.

\bibitem[{\citenamefont{{Knox} and {Song}}(2002)}]{2002PhRvL..89a1303K}
\bibinfo{author}{\bibfnamefont{L.}~\bibnamefont{{Knox}}} \bibnamefont{and}
  \bibinfo{author}{\bibfnamefont{Y.-S.} \bibnamefont{{Song}}},
  \bibinfo{journal}{Physical Review Letters} \textbf{\bibinfo{volume}{89}},
  \bibinfo{pages}{011303} (\bibinfo{year}{2002}), \eprint{astro-ph/0202286}.

\bibitem[{\citenamefont{{Hirata} and
  {Seljak}}(2003{\natexlab{a}})}]{2003PhRvD..67d3001H}
\bibinfo{author}{\bibfnamefont{C.~M.} \bibnamefont{{Hirata}}} \bibnamefont{and}
  \bibinfo{author}{\bibfnamefont{U.}~\bibnamefont{{Seljak}}},
  \bibinfo{journal}{\prd} \textbf{\bibinfo{volume}{67}},
  \bibinfo{pages}{043001} (\bibinfo{year}{2003}{\natexlab{a}}),
  \eprint{astro-ph/0209489}.

\bibitem[{\citenamefont{{Hirata} and
  {Seljak}}(2003{\natexlab{b}})}]{2003PhRvD..68h3002H}
\bibinfo{author}{\bibfnamefont{C.~M.} \bibnamefont{{Hirata}}} \bibnamefont{and}
  \bibinfo{author}{\bibfnamefont{U.}~\bibnamefont{{Seljak}}},
  \bibinfo{journal}{\prd} \textbf{\bibinfo{volume}{68}},
  \bibinfo{pages}{083002} (\bibinfo{year}{2003}{\natexlab{b}}),
  \eprint{astro-ph/0306354}.

\bibitem[{\citenamefont{{Amarie} et~al.}(2005)\citenamefont{{Amarie}, {Hirata},
  and {Seljak}}}]{2005PhRvD..72l3006A}
\bibinfo{author}{\bibfnamefont{M.}~\bibnamefont{{Amarie}}},
  \bibinfo{author}{\bibfnamefont{C.}~\bibnamefont{{Hirata}}}, \bibnamefont{and}
  \bibinfo{author}{\bibfnamefont{U.}~\bibnamefont{{Seljak}}},
  \bibinfo{journal}{\prd} \textbf{\bibinfo{volume}{72}},
  \bibinfo{pages}{123006} (\bibinfo{year}{2005}), \eprint{astro-ph/0508293}.

\bibitem[{\citenamefont{{Verde} et~al.}(2006)\citenamefont{{Verde}, {Peiris},
  and {Jimenez}}}]{2006JCAP...01..019V}
\bibinfo{author}{\bibfnamefont{L.}~\bibnamefont{{Verde}}},
  \bibinfo{author}{\bibfnamefont{H.~V.} \bibnamefont{{Peiris}}},
  \bibnamefont{and}
  \bibinfo{author}{\bibfnamefont{R.}~\bibnamefont{{Jimenez}}},
  \bibinfo{journal}{Journal of Cosmology and Astro-Particle Physics}
  \textbf{\bibinfo{volume}{1}}, \bibinfo{pages}{19} (\bibinfo{year}{2006}),
  \eprint{astro-ph/0506036}.

\bibitem[{\citenamefont{{Sigurdson} and {Cooray}}(2005)}]{2005PhRvL..95u1303S}
\bibinfo{author}{\bibfnamefont{K.}~\bibnamefont{{Sigurdson}}} \bibnamefont{and}
  \bibinfo{author}{\bibfnamefont{A.}~\bibnamefont{{Cooray}}},
  \bibinfo{journal}{Physical Review Letters} \textbf{\bibinfo{volume}{95}},
  \bibinfo{pages}{211303} (\bibinfo{year}{2005}), \eprint{astro-ph/0502549}.

\bibitem[{\citenamefont{{Zaldarriaga} and
  {Seljak}}(1997)}]{1997PhRvD..55.1830Z}
\bibinfo{author}{\bibfnamefont{M.}~\bibnamefont{{Zaldarriaga}}}
  \bibnamefont{and} \bibinfo{author}{\bibfnamefont{U.}~\bibnamefont{{Seljak}}},
  \bibinfo{journal}{\prd} \textbf{\bibinfo{volume}{55}}, \bibinfo{pages}{1830}
  (\bibinfo{year}{1997}), \eprint{astro-ph/9609170}.

\bibitem[{\citenamefont{{Bock} et~al.}(2006)\citenamefont{{Bock}, {Church},
  {Devlin}, {Hinshaw}, {Lange}, {Lee}, {Page}, {Partridge}, {Ruhl}, {Tegmark}
  et~al.}}]{2006astro.ph..4101B}
\bibinfo{author}{\bibfnamefont{J.}~\bibnamefont{{Bock}}},
  \bibinfo{author}{\bibfnamefont{S.}~\bibnamefont{{Church}}},
  \bibinfo{author}{\bibfnamefont{M.}~\bibnamefont{{Devlin}}},
  \bibinfo{author}{\bibfnamefont{G.}~\bibnamefont{{Hinshaw}}},
  \bibinfo{author}{\bibfnamefont{A.}~\bibnamefont{{Lange}}},
  \bibinfo{author}{\bibfnamefont{A.}~\bibnamefont{{Lee}}},
  \bibinfo{author}{\bibfnamefont{L.}~\bibnamefont{{Page}}},
  \bibinfo{author}{\bibfnamefont{B.}~\bibnamefont{{Partridge}}},
  \bibinfo{author}{\bibfnamefont{J.}~\bibnamefont{{Ruhl}}},
  \bibinfo{author}{\bibfnamefont{M.}~\bibnamefont{{Tegmark}}},
  \bibnamefont{et~al.}, \bibinfo{journal}{ArXiv Astrophysics e-prints}
  (\bibinfo{year}{2006}), \eprint{astro-ph/0604101}.

\bibitem[{\citenamefont{{Zaldarriaga} and
  {Seljak}}(1998)}]{1998PhRvD..58b3003Z}
\bibinfo{author}{\bibfnamefont{M.}~\bibnamefont{{Zaldarriaga}}}
  \bibnamefont{and} \bibinfo{author}{\bibfnamefont{U.}~\bibnamefont{{Seljak}}},
  \bibinfo{journal}{\prd} \textbf{\bibinfo{volume}{58}},
  \bibinfo{pages}{023003} (\bibinfo{year}{1998}), \eprint{astro-ph/9803150}.

\bibitem[{\citenamefont{{Lewis} and {Challinor}}(2006)}]{2006PhR...429....1L}
\bibinfo{author}{\bibfnamefont{A.}~\bibnamefont{{Lewis}}} \bibnamefont{and}
  \bibinfo{author}{\bibfnamefont{A.}~\bibnamefont{{Challinor}}},
  \bibinfo{journal}{\physrep} \textbf{\bibinfo{volume}{429}},
  \bibinfo{pages}{1} (\bibinfo{year}{2006}), \eprint{astro-ph/0601594}.

\bibitem[{\citenamefont{{Smail} et~al.}(1995)\citenamefont{{Smail}, {Hogg},
  {Yan}, and {Cohen}}}]{1995ApJ...449L.105S}
\bibinfo{author}{\bibfnamefont{I.}~\bibnamefont{{Smail}}},
  \bibinfo{author}{\bibfnamefont{D.~W.} \bibnamefont{{Hogg}}},
  \bibinfo{author}{\bibfnamefont{L.}~\bibnamefont{{Yan}}}, \bibnamefont{and}
  \bibinfo{author}{\bibfnamefont{J.~G.} \bibnamefont{{Cohen}}},
  \bibinfo{journal}{\apjl} \textbf{\bibinfo{volume}{449}},
  \bibinfo{pages}{L105+} (\bibinfo{year}{1995}),
  \eprint{arXiv:astro-ph/9506095}.

\bibitem[{\citenamefont{{Knox}}(1995)}]{1995PhRvD..52.4307K}
\bibinfo{author}{\bibfnamefont{L.}~\bibnamefont{{Knox}}},
  \bibinfo{journal}{\prd} \textbf{\bibinfo{volume}{52}}, \bibinfo{pages}{4307}
  (\bibinfo{year}{1995}), \eprint{astro-ph/9504054}.

\bibitem[{\citenamefont{{Spergel} et~al.}(2007)\citenamefont{{Spergel}, {Bean},
  {Dor{\'e}}, {Nolta}, {Bennett}, {Dunkley}, {Hinshaw}, {Jarosik}, {Komatsu},
  {Page} et~al.}}]{2007ApJS..170..377S}
\bibinfo{author}{\bibfnamefont{D.~N.} \bibnamefont{{Spergel}}},
  \bibinfo{author}{\bibfnamefont{R.}~\bibnamefont{{Bean}}},
  \bibinfo{author}{\bibfnamefont{O.}~\bibnamefont{{Dor{\'e}}}},
  \bibinfo{author}{\bibfnamefont{M.~R.} \bibnamefont{{Nolta}}},
  \bibinfo{author}{\bibfnamefont{C.~L.} \bibnamefont{{Bennett}}},
  \bibinfo{author}{\bibfnamefont{J.}~\bibnamefont{{Dunkley}}},
  \bibinfo{author}{\bibfnamefont{G.}~\bibnamefont{{Hinshaw}}},
  \bibinfo{author}{\bibfnamefont{N.}~\bibnamefont{{Jarosik}}},
  \bibinfo{author}{\bibfnamefont{E.}~\bibnamefont{{Komatsu}}},
  \bibinfo{author}{\bibfnamefont{L.}~\bibnamefont{{Page}}},
  \bibnamefont{et~al.}, \bibinfo{journal}{\apjs}
  \textbf{\bibinfo{volume}{170}}, \bibinfo{pages}{377} (\bibinfo{year}{2007}),
  \eprint{arXiv:astro-ph/0603449}.

\bibitem[{\citenamefont{{Lewis} et~al.}(2000)\citenamefont{{Lewis},
  {Challinor}, and {Lasenby}}}]{2000ApJ...538..473L}
\bibinfo{author}{\bibfnamefont{A.}~\bibnamefont{{Lewis}}},
  \bibinfo{author}{\bibfnamefont{A.}~\bibnamefont{{Challinor}}},
  \bibnamefont{and}
  \bibinfo{author}{\bibfnamefont{A.}~\bibnamefont{{Lasenby}}},
  \bibinfo{journal}{\apj} \textbf{\bibinfo{volume}{538}}, \bibinfo{pages}{473}
  (\bibinfo{year}{2000}), \eprint{arXiv:astro-ph/9911177}.

\bibitem[{\citenamefont{{Barkana} and {Loeb}}(2007)}]{2007RPPh...70..627B}
\bibinfo{author}{\bibfnamefont{R.}~\bibnamefont{{Barkana}}} \bibnamefont{and}
  \bibinfo{author}{\bibfnamefont{A.}~\bibnamefont{{Loeb}}},
  \bibinfo{journal}{Reports of Progress in Physics}
  \textbf{\bibinfo{volume}{70}}, \bibinfo{pages}{627} (\bibinfo{year}{2007}),
  \eprint{arXiv:astro-ph/0611541}.

\bibitem[{\citenamefont{{Zahn} and {Zaldarriaga}}(2006)}]{2006ApJ...653..922Z}
\bibinfo{author}{\bibfnamefont{O.}~\bibnamefont{{Zahn}}} \bibnamefont{and}
  \bibinfo{author}{\bibfnamefont{M.}~\bibnamefont{{Zaldarriaga}}},
  \bibinfo{journal}{\apj} \textbf{\bibinfo{volume}{653}}, \bibinfo{pages}{922}
  (\bibinfo{year}{2006}), \eprint{astro-ph/0511547}.

\bibitem[{\citenamefont{{Seljak} and {Hirata}}(2004)}]{2004PhRvD..69d3005S}
\bibinfo{author}{\bibfnamefont{U.}~\bibnamefont{{Seljak}}} \bibnamefont{and}
  \bibinfo{author}{\bibfnamefont{C.~M.} \bibnamefont{{Hirata}}},
  \bibinfo{journal}{\prd} \textbf{\bibinfo{volume}{69}},
  \bibinfo{pages}{043005} (\bibinfo{year}{2004}), \eprint{astro-ph/0310163}.

\end{thebibliography}

\end{document}